\documentclass{iopart}

\def\Cas{\mathrm{Cas}}

\def\bk{\mathbf{k}}

\def\dd{\mathrm{d}}

\def\TE{\mathrm{TE}}
\def\TM{\mathrm{TM}}

\def\P{\mathrm{P}}    %plasma
\def\C{\mathrm{C}}    %corrugation or correlation or cavity
\def\PP{\mathrm{PP}}  %plane-plane
\def\lat{\mathrm{lat}}%lateral
\def\GC{G_\C} %corrugation sensitivity function
\def\rC{\rho_\C} %corrugation reduced function
\def\lat{\mathrm{lat}}  
\def\kC{k_\C}    %corrugation
    
\def\PFA{\mathrm{PFA}}
\def\calS{\mathcal{S}}
\def\calR{\mathcal{R}}
\def\calK{\mathcal{K}}
\def\calD{\mathcal{D}}
\def\calF{\mathcal{F}}

\begin{document}

\title[Casimir energy and geometry beyond the PFA]%
{Casimir energy and geometry : beyond the Proximity Force Approximation}

\author{S Reynaud$^1$, PA Maia Neto$^2$ and A Lambrecht$^1$}

\address{$^1$ Laboratoire Kastler Brossel, CNRS, ENS,
Universit\'e Pierre et Marie Curie (UPMC), Campus Jussieu, 75252 Paris, France} 
\address{$^2$ Instituto de F\'{\i}sica,
UFRJ, CP 68528,   Rio de Janeiro,  RJ, 21941-972, Brazil}
\ead{serge.reynaud@spectro.jussieu.fr}

\begin{abstract}
We review the relation between Casimir effect and geometry, emphasizing
deviations from the commonly used Proximity Force Approximation (PFA).
We use to this aim the scattering formalism which is nowadays the best tool
available for accurate and reliable theory-experiment comparisons. 
We first recall the main lines of this formalism when the mirrors 
can be considered to obey specular reflection.
We then discuss the more general case where non planar mirrors give rise
to non-specular reflection with wavevectors and field polarisations mixed.
The general formalism has already been fruitfully used for evaluating the 
effect of roughness on the Casimir force as well as the lateral Casimir force 
or Casimir torque appearing between corrugated surfaces.
In this short review, we focus our attention on the case of the lateral force 
which should make possible in the future an experimental demonstration of 
the nontrivial ({\it ie} beyond PFA) interplay of geometry and Casimir effect.

\end{abstract}

\submitto{\JPA, QFEXT'07 special issue}
\date{\today}
%\maketitle

\section{Introduction}

The Casimir force \cite{Casimir} is a remarkable prediction of
quantum field theory.
As the most easily accessible effect of vacuum fluctuations
in the macroscopic world, it deserves careful experimental tests
\cite{Milonni94,LamoreauxResource99,Reynaud01}.
After tests which confirmed its existence and main properties \cite{Sparnaay89},
experiments have been largely improved by technological achievements mastered 
over the last decade \cite{Lamoreaux97,Mohideen98,Harris00,Ederth00,%
Bressi02,Decca03prl,ChenPRA04,DeccaAP05}.
Meanwhile, it was realized that the Casimir force, a dominant force 
at micron or sub-micron distances, was clearly an important aspect
of the study of micro- and nano-oscillators (MEMS, NEMS) 
\cite{Roukes01,Chan01}. 

These recent advances have been reviewed in a number of papers, for example
\cite{Bordag01,LambrechtPoincare,Milton05} and in a special issue of the 
New Journal of Physics \cite{NJP06}.
In the next paragraphs, we emphasize arguments which plead for careful 
comparisons between experimental measurements and theoretical
predictions of the Casimir force \cite{LambrechtNJP06}.

\subsection*{Why testing the Casimir force}

A precise knowledge of the Casimir force is a key point for the tests
of gravity at sub-millimeter ranges \cite{Fischbach98,Adelberger03,OnofrioNJP06}.
A strong constraint has been obtained recently in short range
Cavendish-like experiments \cite{Kapner07}.
Should an hypothetical new force have a Yukawa-like form, its strength
could not be larger than that of gravity for Yukawa ranges larger than
56 $\mu$m.

Tests performed at shorter ranges essentially amount to comparisons 
with theory of Casimir force measurements. In other words, the 
looked for hypothetical new force would correspond to an observable
given by the difference $F_\mathrm{exp}-F_\mathrm{th}$ between
experiment and theory. This implies that the theoretical prediction
$F_\mathrm{th}$ and the experimental measurement $F_\mathrm{exp}$ 
have to be treated independently from each other and 
with the same accuracy and reliability requirements.

To sum up the argument, the fact that Casimir force experiments could
be a window on hypothetical deviations from standard physics forbids
one to use theory-experiment comparison as an argument for proving 
(or disproving) some specific experiment or theoretical model.
In this context, it is important to use a theoretical formalism 
having the ability to take into account the significant differences 
between the real experimental conditions and the
ideal situation studied by Casimir
\cite{Reynaud01,LambrechtPoincare,LambrechtNJP06}.

Casimir calculated the force between a pair of perfectly smooth, flat and
parallel plates in the limit of zero temperature and perfect reflection.
He found expressions for the force $F_\Cas$ and energy $E_\Cas$ which only depend
on the distance $L$, the area $A$ and two fundamental constants, 
the speed of light $c$ and Planck constant $\hbar$
\begin{eqnarray}
F_\Cas = \frac{\hbar c \pi ^2 A}{240L^4} =
\frac{\dd E_\Cas}{\dd L} \quad, \quad
E_\Cas = - \frac{\hbar c \pi^2 A}{720 L^3} 
\label{Fcasimir}
\end{eqnarray}
This universality property of the Casimir expression is related to the assumption
of perfect reflection that is also to the saturation of the optical response 
of the mirrors when they reflect 100\% of the incoming light. 
However, no real mirror can be considered as a perfect reflector at all field
frequencies. The most precise experiments are performed with
metallic mirrors which are good reflectors only at frequencies smaller
than their plasma frequency. It follows that the Casimir 
force can obey the Casimir expression only at distances $L$ 
larger than the plasma wavelength $\lambda _\P$.

As this effect of imperfect reflection is large,
a precise knowledge of its frequency dependence is essential for obtaining an
accurate theoretical prediction of the Casimir force \cite{LambrechtEPJ00}.
This is also true for another correction to the ideal Casimir formula
associated with temperature effect. For discussions of this effect,
we refer to discussions in \cite{GenetPRA00,Reynaud03} and the recent 
review \cite{BrevikNJP06}. 
We focus now our attention on the effects of geometry which are
also important in this context.

\subsection*{Why testing the effects of geometry}

It has been repeatedly stated over the years that the connection
between the Casimir effect and geometry should show a rich variety
of sensitive dependences \cite{Balian7778,Plunien86,Balian0304}.
The basis for this statement is the important fact that the 
Casimir forces cannot be additive, except in the specific case
of interaction between very dilute media.

Meanwhile, most experiments are performed between a plane and a sphere 
with the Casimir force in this geometry calculated using the
Proximity Force Approximation \cite{Derjaguin68}, though
the latter amounts to a mere averaging over the distribution
of local inter-plate distances.
The PFA is expected to be valid in the plane-sphere geometry when the 
radius $R$ is much larger than the separation $L$ \cite{Schaden00,Jaffe04}
and it is used to analyze most present day experiments.
Results going beyond this approximation have been obtained recently 
\cite{Gies03,JaffePRA05,Bulgac06,Bordag06,Emig06,Dalvit06,Emig07,Milton07,%
Krause07,Rodriguez07}.
Some of these theoretical models involve scalar fields reflected on 
perfect boundary conditions and can hardly be compared with
experiments but those dealing with electromagnetic fields 
have now to be used for comparisons with experimental results 
obtained in the plane-sphere geometry.

As the effects of geometry on the Casimir force open access 
to a rich and stimulating physics, it is also important to 
explore this domain through new dedicated experiments. 
Only a few experiments have been designed to this aim 
which use a specific geometry with
periodic corrugations imprinted on metallic surfaces.
In this case, the Casimir force contains a lateral component 
since lateral translation symmetry is broken \cite{Golestanian}. 
The lateral Casimir force is smaller than the normal one, but it has
nevertheless already been measured in experiments \cite{Chen02}. 
The results have been found to agree within a bar of $\pm24\%$ with 
PFA calculations.

Calculations beyond the PFA have also been performed by using more 
elaborate theoretical methods. 
The lateral force has been evaluated for perfectly reflecting mirrors 
using a path-integral formulation in a perturbative \cite{Emig0103} 
or non perturbative approach \cite{Emig05}. 
As the experiments are performed at distances $L$ not much larger
than the plasma wavelength $\lambda_\P$, it is essential to account for 
the optical properties of the metals \cite{LambrechtEPJ00,GenetPRA00}.
Below we will present results obtained for corrugated metallic mirrors
in the limiting case where the corrugation can be treated as a
small perturbation \cite{Maia05,Rodrigues06,Rodrigues07}.
As expected, the PFA is found to be valid 
when the corrugated surfaces appear as nearly plane to the vacuum fields
involved in the calculation of the Casimir energy,
that is to say when the corrugation wavelength $\lambda_\C$ 
is larger than the other relevant length scales. 

\subsection*{Outline of the paper}

We review below the theory of Casimir effect within scattering theory.
We will in particular present the formula giving the (QED) theoretical
prediction for the Casimir force between scatterers placed in vacuum,
or more generally at thermodynamical equilibrium with $T\neq0$.
This formula has been written years ago for plane and parallel 
mirrors showing specular reflection \cite{Jaekel91}.
It has been used to discuss in a qualitative manner the effect
of reflection properties of the mirrors on the Casimir force
\cite{LambrechtPoincare,LambrechtEPJ00,GenetPRA00,GenetPRA03}.
Its applicability domain has been enlarged up to the point where it is 
now capable of dealing with non planar geometries with non specular 
reflection mixing field polarizations and transverse wave-vectors
\cite{LambrechtNJP06}.
We will recall below the application of this method to the calculation
of the lateral Casimir force between corrugated plates 
\cite{Rodrigues06,Rodrigues07}.

Note that similar discussions have been devoted to the discussion of the 
effect of surface roughness on the Casimir force. 
This description is commonly given within the PFA \cite{Klimchitskaya99} 
which cannot remain valid for arbitrary roughness wavelengths \cite{Genet03}.
As the effect of roughness is only a small correction of the Casimir force,
one can however hardly expect quantitative theory-experiment
comparisons in this case. This is why we will not discuss it below.
Other applications have also been presented for the Casimir torque
appearing between misaligned corrugation plates \cite{RodriguesEPL06}
and for the Casimir-Polder force between an atom or a cloud of atoms (BEC)
and a corrugated metallic plate \cite{Dalvit07}.

\section{Specular scattering}

We first consider the geometry with perfectly plane and parallel mirrors
aligned along the directions $x$ and $y$.
As the configuration obeys a symmetry with respect to time translation
as well as lateral space translations (along directions $x$ and $y$),
the frequency $\omega$, transverse vector $\bk \equiv \left( k_{x},k_{y}\right)$ and
polarization $p=\TE,\TM$ are preserved by the scattering processes which
couple field modes having the same values for the preserved quantum
numbers but a different sign for the longitudinal wavevector $k_z$.
The two mirrors $j=1,2$ are described by reflection and transmission amplitudes 
which depend on frequency, incidence angle 
$\theta=\arccos\left(ck_z/\omega\right)$ and polarization $p$.

\subsection*{Scattering formulas}

The important result is that the Casimir force can be written in terms of the reflection 
amplitudes $r_j$ of the two mirrors, as seen from inside the Fabry-Perot cavity 
formed by the two mirrors \cite{Jaekel91}.
In order to write this relation, we introduce two functions which characterize 
the optical response of the cavity to an input field (dependences with respect to
$\omega$, $\bk$ and $p$ are omitted)
\begin{eqnarray}
&&f = \frac{r_1 r_2 e^{2ik_z L}}{1-r_1 r_2 e^{2ik_z L}} \quad,\quad
g = 1 + f + f^* =
\frac{1-\left|r_1 r_2 e^{2ik_z L}\right| ^2}{\left|1-r_1 r_2 e^{2ik_z L}\right| ^2}
\label{fg}
\end{eqnarray}
$f$ is the closed-loop function describing the cavity
($L$ is the length of the cavity) and, therefore, obeys analyticity properties. 
Meanwhile $g$ is the ratio of energy inside the cavity to energy outside the cavity,
that is also the ratio of spectral density inside the cavity 
to spectral density outside the cavity for a given mode.
Its expression is valid for lossy as well as lossless mirrors
as was demonstrated with an increasing range of validity
in \cite{Jaekel91}, \cite{Barnett98} and \cite{GenetPRA03}. 
For lossy mirrors, it accounts for the additional
fluctuations accompanying losses inside the mirrors.

Assuming thermal equilibrium for the whole ``cavity + fields'' system, 
we obtain the radiation pressure exerted by the field fluctuations
upon the mirrors. This leads to the following expression of the
Casimir force as the sum over all field modes $m$ of this
radiation pressure ($m$ gathers the parameters $\omega$, $\bk$ and $p$)
\begin{eqnarray}
F &=&  
\sum_m \left( \frac 12 + \overline{n} \right) \hbar\omega 
\cos^2\theta \left\{1-g\right\} \nonumber\\
&=& - \sum_m \left( \frac 12 + \overline{n} \right) \hbar\omega 
\cos^2\theta \left\{f+f^*\right\}
\label{ForceReal}
\end{eqnarray}
Here $\left( \frac 12 + \overline{n} \right) \hbar\omega$
is the mean energy per mode at temperature $T$ with $\overline{n}$ 
the mean number of photons per mode ($\overline{n}=0$ at $T=0$, $\overline{n}>0$ 
otherwise) 
\begin{eqnarray}
\frac 12 + \overline{n} &=& \frac 12 \coth
\frac {\hbar\omega}{2k_\mathrm{B}T}
\end{eqnarray}
Meanwhile $\cos^2\theta$ is a projection factor appearing in the translation
from energy density to pressure; finally $\left\{1-g\right\}$
represents the difference between pressures on the outer and inner
sides of the mirrors respectively.
Equation (\ref{ForceReal}) contains the contribution of ordinary
modes freely propagating outside and inside the cavity 
($\omega > c |\bk|$), which merely reflects the intuitive picture of 
a radiation pressure of field fluctuations on the mirrors \cite{Jaekel91}.
But it also includes the contribution of evanescent waves ($\omega < c |\bk|$)
which propagate inside the mirrors with an incidence angle larger than the 
limit angle \cite{GenetPRA03}. The properties of the latter are described 
through an analytical continuation of those of ordinary waves, using the well 
defined analytic behavior of the function $f$.

Equation (\ref{ForceReal}) can be equivalently written as a differential with
respect to length of a free energy
\begin{eqnarray}
F=\frac{\partial\calF}{\partial L} \quad,\quad
\calF &=&  \frac{\hbar c}i
\sum_m \left( \frac 12 + \overline{n} \right) 
\ln \frac{1-r_1 r_2 e^{2ik_z L}}{1-r_1^* r_2^* e^{-2ik_z L}}
\label{EReal}
\end{eqnarray}
Using analyticity properties, as well as high frequency transparency to
neglect the contribution of large frequencies, both equations (\ref{ForceReal}) 
and (\ref{EReal}) can be transformed into integral over imaginary frequencies
$\omega=i\xi$.
We will write below more general forms of these relations, valid also
for non specular scattering.

\subsection*{The Lifshitz formula as a particular case}

Equations (\ref{ForceReal}) and (\ref{EReal}) reproduce the Casimir formulas
(\ref{Fcasimir}) in the limits of perfect reflection $r_1 r_2 \rightarrow 1$
and null temperature $T \rightarrow 0$. They are regular for any optical model 
of mirrors obeying causality and high frequency transparency properties,
without needing any further regularization.
They can thus be used for calculating the Casimir force between arbitrary
mirrors, as soon as the reflection amplitudes are specified. 
These amplitudes are commonly deduced from microscopic models of mirrors,
the simplest of which is the well known Lifshitz model \cite{Lifshitz56}.

This model corresponds to plates having a large optical thickness,
and characterized by a local dielectric function 
$\varepsilon \left( \omega\right)$.
The reflection amplitudes are thus given by the Fresnel law
written at the vacuum-bulk interface 
\begin{eqnarray}
&&r^\TE = \frac{k_z-K_z}{k_z+K_z} \quad,\quad
r^\TM = \frac{K_z-\varepsilon k_z}{K_z+\varepsilon k_z} 
\nonumber \\
&&ck_z = \sqrt{\omega ^2-c^2 k_z^2}  \quad,\quad
cK_z = \sqrt{ \varepsilon \omega ^2 - c^2 k_z^2} 
\label{rThick}
\end{eqnarray}
$k_z$ and $K_z$ correspond to the longitudinal wavevector in vacuum and
in the bulk respectively.
Taken with equations (\ref{ForceReal}) and (\ref{EReal}) (possibly
translated to the domain of imaginary frequencies), these relations 
(\ref{rThick}) reproduce the Lifshitz expression for the Casimir force
\cite{Lifshitz56}. The latter tend to the original Casimir expression
in the limit $\varepsilon \rightarrow \infty$ which produces perfectly
reflecting mirrors \cite{Schwinger78}. 

At this stage, several remarks are worth being emphasized~:
\begin{itemize}
\item The expression of the force was not written in this
manner by Lifshitz. To our best knowledge, Kats \cite{Kats77} was
the first to notice that Lifshitz expression could be written in
terms of the reflection amplitudes. 

\item The Lifshitz expression is valid for the cases for which it was 
derived. Its extension to more general situations can only be considered
as valid after a careful examination of the derivation.

\item In the most general case, the optical response of the bulk
material cannot be described by a local dielectric function.
In this case, the description in terms of reflection amplitudes, 
which necessarily differ from the specific expressions (\ref{rThick}), 
is still valid \cite{Jaekel91,GenetPRA03,LambrechtNJP06}. 

\end{itemize}

\subsection*{Description of real mirrors}

In order to obtain a quantitative description of the effect of finite conductivity,
we may in a first approach use the expressions (\ref{rThick}) with the dielectric
function corresponding to the plasma model ($\omega _\P$ the plasma frequency)
\begin{eqnarray}
\varepsilon \left( \omega \right) =1-\frac{\omega _\P ^{2}}{\omega ^{2}} \quad,\quad
\varepsilon \left( i\xi \right) =1+\frac{\omega _\P ^{2}}{\xi ^{2}}
\label{epsPlasma}
\end{eqnarray}
When performing these calculations, one recovers as expected
the Casimir formula at large distances
($F\rightarrow F_\Cas$ when $L\gg\lambda_\P$). 
At distances smaller
than $\lambda _\P $ in contrast, a significant reduction is obtained with
the asymptotic law of variation read as 
\begin{eqnarray}
L &\ll &\lambda _\P \quad \rightarrow \quad \frac{F}{F_\Cas} \simeq 1.193  \frac{L}{\lambda _\P }  
\end{eqnarray}
This can be understood as the result of the Coulomb interaction of
surface plasmons at the two vacuum/metal interfaces
\cite{GenetAFLdB04,Henkel04}. The generalization of this idea at
arbitrary distances is more subtle since it involves a full
electromagnetic treatment of the plasmon as well as ordinary photon
modes \cite{plasmon}.

The plasma model cannot provide a fully satisfactory description of the
optical response of metals.
A more realistic representation of the metals includes the description
of the relaxation processes of conduction electrons
as well as that of interband transitions. 
The reader is referred to \cite{LambrechtEPJ00} for a more detailed
discussion \cite{data}.
The values of the complex index of refraction for different metals, 
measured through different optical techniques, are tabulated in several
handbooks \cite{handbooks}. Optical data may vary from one
reference to another, leading to different estimations of the Casimir 
force \cite{Pirozhenko06}.
Let us emphasize that the problem here is neither due to a lack of precision
of the calculations nor to inaccuracies in experiments.
The problem is that calculations and experiments may consider
physical samples with different optical properties.
This difficulty should be solved by measuring the reflection amplitudes
of the mirrors used in the experiment and inserting these informations
in the formula giving the predicted Casimir force.

\subsection*{Temperature correction}

The Casimir force between metallic mirrors at non zero temperature
has given rise to contradictory claims which have raised doubts about
the theoretical expression of the force. We do not repeat here the
discussions which have been devoted to the topic in 
\cite{GenetPRA00,Reynaud03,LambrechtNJP06} (see also the recent 
review \cite{BrevikNJP06} and contributions on the topic 
in the present volume \cite{thermal}). 
We only want to stress again that the running controversy can only 
be solved through an improvement of the knowledge of the reflection 
amplitudes, particularly at low frequencies. 
As already discussed, the best manner to do that is to measure these
amplitudes on the mirrors used in the experiment

\section{Non specular scattering}

We will now present a more general formalism where the Casimir force
and energy are calculated between two objects with non planar shapes.
This formalism is an extension of what has already been presented
with the scattering amplitudes now accounting for non-specular reflection.
The non-specular case is of course the generic one
while specular reflection can only be an idealization.
After an introduction to this general formalism, we will discuss
applications to the lateral force between corrugated mirrors and
we will in particular emphasize deviations from the PFA.

\subsection*{General scattering formulas}

In order to introduce the general formalism, let us first rewrite
expression (\ref{EReal}) of the Casimir free energy between two 
parallel plane plates as the sum over modes 
\begin{eqnarray}
&&F=\frac{\partial\calF}{\partial L} \quad,\quad
\calF = i\hbar \int_0^\infty \frac{\dd \omega}{2\pi}
\left( \frac 12 + \overline{n} \right) 
\ln \det \calS \nonumber\\
&&\ln \det \calS = \Tr \ln \calS = \Tr \ln \frac{d^*}{d}
\quad,\quad d\equiv 1-r_1 r_2 e^{2ik_z L}
\label{phaseshift}
\end{eqnarray}
These equations correspond to the following interpretation \cite{Jaekel91}~:
The force $F$ is the change of the free energy $\calF$ when the scatterers are being displaced.
The free energy $\calF$ is described by a storage of vacuum energy due to the scattering process,
and it is written in terms of the $\calS$-matrix associated with the cavity.
As the scattering on stationary objects preserves frequency, this $\calS$-matrix is defined at 
a given value of $\omega$. As the surfaces are plane and parallel, the scattering also preserves
the transverse wavevector $\bk$ and polarization $p$ 
(it only couples modes with opposite values of the longitudinal wavevector).
The symbol $\Tr$ in (\ref{phaseshift}) refers to a trace over the modes corresponding to different
values of $\bk$ and $p$ at a fixed frequency.
The quantity $\ln \det \calS$ can be written in terms of the matrix $d$ which is diagonal
on the basis of plane waves, so that equation (\ref{phaseshift}) is effectively equivalent
to (\ref{EReal}).
This ``scattering formula'' or ``phaseshift formula'' \cite{Jaekel91} can
equivalently be written as a sum over imaginary frequencies $\omega=i\xi$
\begin{eqnarray}
\label{phaseshiftIm} 
&&\calF = \hbar \int_0^\infty \frac{\dd \xi}{2\pi}
\left( 1+ 2\overline{n} \right) \ln \det d  \\
&&d\equiv 1-r_1 r_2 \exp\left(-2\sqrt{{\bk^2+\xi^2}} L\right)
\nonumber
\end{eqnarray}
$d$ is the denominator of the loop function (\ref{fg})
here written for imaginary frequencies.

As a consequence of this interpretation, it is clear that 
a more general formula of the Casimir energy can be written
in a similar manner for the case of stationary but non-specular scattering
\cite{Maia05,LambrechtNJP06}.
It can be expressed either as a sum over real frequencies, including 
ordinary and evanescent waves, or as a sum over imaginary frequencies
\begin{eqnarray}
\label{Fnonspec}
&&\calF = \hbar \int_0^\infty \frac{\dd \xi}{2\pi}
\left( 1+ 2\overline{n} \right) \ln \det \calD \\
&&\calD \equiv 1-\calR_1 \exp\left(-\calK L\right) \calR_2 \exp\left(-\calK L\right) 
\nonumber
\end{eqnarray}
The matrices $\calD$, $\calR_1$ and $\calR_2$ are no
longer diagonal on the basis of plane waves since they describe non specular
reflection on the two mirrors. The propagation factors contained in $\calK$ 
remain diagonal on the basis of plane waves with their diagonal values written
as in (\ref{phaseshiftIm}). Clearly the expression (\ref{Fnonspec}) does 
not depend on the choice of this specific basis. 
Note that the matrices in (\ref{Fnonspec}) do not commute with each other. 
In particular, the two propagation matrices $\exp\left(-\calK L\right)$ appearing in 
$\calD$ can be moved through circular permutations in the product but not adjoined 
to each other.

This equation takes a simpler form at the limit of null temperature
(note the change of notation from the free energy $\calF$ to the 
ordinary energy $E$)
\begin{eqnarray}
&&F=\frac{\dd E}{\dd L} \quad,\quad
E = \hbar \int_0^\infty \frac{\dd \xi}{2\pi} \ln \det \calD 
\label{Enonspec}
\end{eqnarray}
Formula (\ref{Enonspec}) has already been used to evaluate the effect of roughness 
\cite{Maia05} or corrugation \cite{Rodrigues06,Rodrigues07} of the mirrors.
To this aim, it was dealt with in a perturbative manner at second order in the 
roughness or corrugation amplitudes, recalled in the forthcoming paragraphs.
It is clear that it has a larger domain of application, not limited to the 
perturbative regime, as soon as some technique is available for exploiting 
its general form for specific problems of physical interest.
Such a technique has been developed recently by Emig, Graham, Jaffe and Kardar
\cite{Emig07}, through a multipole expansion well adapted to the treatment
of ``compact'' objects, typically two spheres not too close to each other.
The general formula used as the starting point of the expansion is
equivalent to our formula (\ref{Enonspec}) with $\calD$ given in (\ref{Fnonspec}). 
In particular, the $T-$matrices in \cite{Emig07} are identified as the 
non-specular reflection matrices $\calR$ of \cite{Maia05}.
Meanwhile the $U-$matrices in \cite{Emig07} correspond to the propagation matrices
$\exp\left(-\calK L\right)$ of \cite{Maia05}, the difference in their explicit 
expression arising from the fact that they are written in different basis.

\subsection*{Scattering formula for the lateral Casimir force}

We now come to the discussion of the effect of non-planar geometries
and particularly of the deviation from the PFA which could be seen
in experiments. As already stated, we thus focus our attention
on the lateral Casimir force appearing between corrugated plates.
In this case, the deviation of PFA should indeed be visible as
a factor in front of the whole effect. This situation is clearly more
favorable to theory/experiment comparison than that met when 
studying the roughness correction to the normal Casimir force,
with this correction being only a small part of the force \cite{Maia05}.
Stated differently, the lateral Casimir force could allow for a new 
test of a prediction of Quantum ElectroDynamics, namely the dependence
with respect to corrugation wavevector discussed below 
\cite{Rodrigues06,Rodrigues07}.

Here, we consider two parallel plane mirrors, M1 and M2, with corrugated surfaces
described by uniaxial sinusoidal profiles (see Fig.~1 in \cite{Rodrigues07})
\begin{eqnarray}
\label{h1h2}
&&h_1=a_1\,\cos(\kC x) \quad,\quad h_2=a_2\,\cos\left(\kC (x-b)\right) 
\quad,\quad \kC=\frac{2\pi}{\lambda_\C}
\end{eqnarray}
The functions $h_1(x,y)$ and $h_2(x,y)$ measure the local height with respect 
to mean planes $z_1=0$ and $z_2=L$. 
They are defined so that $h_1$ and $h_2$ have null spatial averages, 
$L$ thus representing the mean distance between the two surfaces;
$h_1$ and $h_2$ are both counted as positive when they correspond to 
separation decreases;
$\lambda_\C$ is the corrugation wavelength, $\kC$ the corresponding wave vector,
and $b$ the spatial mismatch between the corrugation crests.

In the following, we will suppose that the corrugation amplitudes are smaller 
than the other length scales, namely the corrugation wavelength $\lambda_\C$,  
the plasma wavelength $\lambda_\P$ and the interplate distance $L$ 
\begin{equation}
\label{perturbat}
a_1, a_2 \ll \lambda_\C, \lambda_\P, L
\end{equation}
Using the PFA, the Casimir energy is thus obtained by adding 
the contributions of various surface elements calculated for distributed
local distances. 
Using the condition (\ref{perturbat}) and expanding up to second order 
in the corrugation amplitudes, we find the lowest-order 
correction to energy within the PFA
\begin{eqnarray} 
\label{PFA2}
\delta E_\PFA &=&  
\frac 12 \frac{\partial^2 E_\PP}{\partial L^2} \; 
\left( \frac{a_1^2+a_2^2}2 + a_1 a_2 \cos(\kC b) \right) 
\end{eqnarray}
with $E_\PP$ the energy calculated between two parallel plane plates.
As the energy corrections proportional to $a_1^2$ and $a_2^2$ 
do not depend on the lateral mismatch $b$, they do not contribute 
to the lateral force which is simply read as
\begin{equation}
\label{FlatPFA}
F_\PFA^\lat = -\frac{\partial \delta E_\PFA}{\partial b} = 
\frac 12 \frac{\partial^2 E_\PP}{\partial L^2} \; \kC a_1 a_2
\sin(\kC b) 
\end{equation}
We will now write the scattering formula for the lateral Casimir force, 
in a perturbative expansion with respect to the corrugation amplitudes.
As in (\ref{FlatPFA}), the correction of the Casimir energy will arise 
at second order in the corrugation amplitudes, with crossed terms of the form 
$a_1\,a_2$ which have the ability to induce lateral forces.
The main difference with (\ref{PFA2},\ref{FlatPFA}) will be the appearance 
of a more complicated dependence in the corrugation wavevector $\kC$.

For this purpose, we expand the non-specular reflection matrix $\calR_j$ 
as the sum $\calR_j^{(0)} + \delta\calR_j$ of a zero-th order contribution 
identified as the specular reflection and of a first-order contribution 
induced by the reflection on the corrugation~\cite{Maia05}.
The lowest order modification of the Casimir energy (\ref{Enonspec}) 
able to produce a lateral force (cross terms $\propto a_1a_2$) is thus read as
\begin{eqnarray}
\delta E = 
- \hbar \int_0^{\infty} \frac{\dd\xi}{2\pi}\,
\Tr \left( \frac{\exp\left(-\calK L\right)}{\calD^{(0)}} \delta\calR_1 
\frac{\exp\left(-\calK L\right)}{\calD^{(0)}} \delta\calR_2  \right)
\label{Ec}
\end{eqnarray}
$\calD^{(0)}$ is the matrix $\calD$ evaluated at zeroth order in the
corrugation. It is diagonal on the basis of plane waves and therefore
commutes with $\calK$.

\subsection*{Explicit results for the plasma model}

In order to obtain explicit expressions, it is then necessary to use some
microscopic model. To this aim, we study the case of bulk metallic plates 
described by the plasma dielectric function.
The non specular reflection amplitudes are then calculated in the Rayleigh
approximation using techniques which have been developed for treating
scattering on rough plates \cite{Agarwal77,Greffet88}.
We want to emphasize that this microscopic model allows one to calculate 
the lateral Casimir force for arbitrary relative values of the three 
parameters $\lambda_\P$, $\lambda_\C$ and $L$, the corrugation
amplitudes remaining the smallest length scale for perturbation
theory to hold (see conditions \ref{perturbat}). 

This calculation leads to the following expression of the lateral Casimir force 
\begin{eqnarray}
F^\lat = -\frac{\partial \delta E}{\partial b} \quad,\quad 
\delta E = \frac A2 \GC(\kC) a_1 a_2 \cos (\kC b) 
\label{Epp1/2}
\end{eqnarray}
with the function $\GC(\kC)$ calculated in \cite{Rodrigues07}.
It is worth emphasizing that the PFA is recovered in equation (\ref{Epp1/2}) as
the limiting case $\kC\rightarrow0$, that is also for long corrugation wavelengths.
This follows from a properly formulated ``Proximity Force Theorem''
\begin{equation}
\lim_{\kC \to 0}A\GC(\kC)= \frac{\dd^2 E_\PP}{\dd L^2}
\label{PFA3}
\end{equation}
This property is ensured, for any model of the material medium, 
by the fact that $\GC(\kC \to 0)$ is given by the specular limit of non 
specular reflection amplitudes~\cite{Rodrigues07}.
This theorem has to be distinguished from the approximation (PFA)
which consists in an identification between $\GC(\kC)$ and its limiting
value $\GC(0)$.
For arbitrary values of $\kC$, the deviation from the PFA is 
described by the ratio
\begin{equation}
\rC(\kC)=\frac{\GC(\kC)}{\GC(0)}
\label{PFA4}
\end{equation}
The variation of this ratio $\rC$ with the various parameters
has been described in a detailed manner in \cite{Rodrigues06,Rodrigues07}.
Some curves are drawn as examples in the Fig.~1 of \cite{Rodrigues06}
with $\lambda_\P=137$nm chosen to fit the case of gold covered plates. 
An important feature is that $\rC$ is smaller than unity as soon as
$\kC$ significantly deviates from 0.
For large values of $\kC$, it even decays exponentially to zero.

\section{Concluding remarks}

We have studied the lateral Casimir force between two corrugated metallic plates.
To this aim, we have used a general scattering formula in a perturbative regime
corresponding to corrugation amplitudes smaller than the other length scales 
$L$, $\lambda_\C$ and $\lambda_\P$. 
The result describes a variety of situations where these three scales have
arbitrary relative values.
The results known for perfect mirrors \cite{Emig05} are recovered
when $\lambda_\P\ll\lambda_\C,L$. 
The Proximity Force Approximation (PFA) is recovered at the limit
of smooth plates $L,\lambda_\P\ll\lambda_\C$.
A third limiting case has been studied in \cite{Rodrigues07}
which corresponds to the opposite case
of rugged corrugations $\lambda_\C \ll L, \lambda_\P$.
This case corresponds to evaluations far beyond the PFA regime 
and is particularly interesting as it constitutes 
a non trivial interplay between
geometry and the Casimir effect \cite{Balian0304}.
It is also of great interest for applications to surfaces with 
structurations at the nanometric scale. 

The numerical figures presented in \cite{Rodrigues06,Rodrigues07}
suggest that non trivial effects of geometry, \textit{i.e.}
effects beyond the PFA, could be observed with dedicated 
lateral force experiments.
Existing experiments by Chen \textit{et al} \cite{Chen02} have used
large corrugation amplitudes $a_1, a_2$ in order to increase the
magnitude of the force. As they do not meet the conditions of validity 
of our perturbative expansion, it is not possible to compare directly
the experimental and theoretical results.
Chen \textit{et al} have found their measurements to agree with PFA 
to within $\pm 24\%$.
Considering smaller amplitudes $a_1, a_2$ with the same values for
the parameters $L$, $\lambda_\C$ and $\lambda_\P$, we have obtained 
a deviation from the PFA of the order of $40\%$, which means that
these parameters do not belong to the domain of validity of the PFA,
at least at the perturbative limit.
 
More work is clearly needed in order to settle this potential concern
in the theory-experiment comparison 
\cite{CommentPRL2007,ReplyPRL2007,Rodrigues07}. 
Progress on this question could be achieved by calculating higher order 
corrections for metallic mirrors beyond the PFA.
These corrections would affect the theoretical predictions, but it
seems unlikely that they would compensate exactly the deviation from PFA 
which has been obtained in the perturbative theory. 
Progress could alternatively come from experiments with
smaller corrugation amplitudes, allowing for a direct comparison
with the perturbative theory.
A better experimental accuracy would also be very valuable,
allowing one to distinguish more easily between alternative
predictions. 
Of course, this program raises serious experimental challenges, 
given the minuteness of the lateral force effect. 
But the reward would be remarkable with potentially the first experimental
demonstration of a nontrivial interplay between geometry and the Casimir effect.

\ack 
The authors thank M.T. Jaekel, C. Genet, R. Rodrigues and D.A.R. Dalvit
for stimulating discussions.
PAMN acknowledges partial finantial support from FAPERJ, CNPq
and Institutos do Mil\^enio de Informa\c c\~ao Qu\^antica e Nanoci\^encias.
AL acknowledges partial financial support 
from the European Contract STRP 12142 NANOCASE.

\end{document}